\documentclass[rnote]{aa}
\usepackage{hyperref}
\usepackage{graphicx}
\usepackage{amssymb}
\usepackage{epstopdf}
\usepackage{times}
\usepackage{natbib}
\newcommand{\D}{\partial}

\newcommand{\TwoLines}[2]{%
  \includegraphics[width=.47\textwidth]{#1}&
  \includegraphics[width=.47\textwidth]{#2}\\
}

\begin{document}

\title{A simple model for electron plasma heating in supernova remnants. }

\author{D. Malyshev \inst{1}
\and S. Gabici \inst{2,1}
\and L. O'C. Drury \inst{1}
\and F. A.  Aharonian \inst{1,3}}

\institute{Dublin Institute for Advanced Studies, 31 Fitzwilliam Place, Dublin 2, Ireland; \and Laboratoire APC, CNRS-UMR 7164, Universit{\'e} Paris 7, 10, rue A. Domon et L{\'e}onie Duquet, 75025 Paris Cedex 13, France; \and Max-Planck-Institut f\"ur Kernphysik, Saupfercheckweg 1, 69117 Heidelberg, Germany
}

\date{}

\abstract{Multiwavelength observations of supernova remnants can be explained within the framework of
diffusive shock acceleration theory,  which allows effective conversion of the explosion energy
into cosmic rays.  Although the models of nonlinear shocks describe reasonably well
the nonthermal  component of emission,  certain issues, including the heating of the thermal
electron plasma and the related X-ray emission, still remain open.}
{Study of electron heating in supernova remnants in the Sedov phase due to Coulomb exchange between protons and electrons.}
{Numerical solution of the equations of the Chevalier model for supernova remnant evolution, coupled with Coulomb scattering heating of the electrons.}{The electron temperature and the X-ray thermal Bremsstrahlung emission from supernova remnants have been calculated as functions of the relevant parameters. Since only the Coulomb mechanism was considered for electron heating, the values obtained for the electron temperatures should be treated as lower limits. Results from this work can be useful to constrain model parameters for observed SNRs.}{}


\maketitle

\section{Introduction}
Multiwavelength observations of supernova remnants (SNRs) allow one, in certain cases, to determine the temperatures $T_e$ and $T_p$ of both electrons and protons behind the shock, \citep[see][for a recent review]{rakowski2005}. The ratio $T_e/T_p$ of these temperatures varies significantly, from $\lesssim 0.07$ for SN 1006 (\citealt{Laming1996}, \citealt{Ghavamian2002}) to $0.67-1$ in the Cygnus Loop \citep{Ghavamian2001}. From a theoretical point of view there are two limits for this ratio. The lower limit, given by the electron to proton mass ratio $T_e /T_p=m_e/m_p$, is the so called total non-equilibrium case, which corresponds to equal thermal velocities for electrons and protons; the upper limit, $T_e=T_p$ is the total equilibrium case, which corresponds to equal energies of electron and proton plasmas.  As was shown by \cite{Ghavamian2001, Ghavamian2002} for the SNRs Cygnus Loop, RCW 86, Tyco and SN 1006, observed values lie between these two limits.
A compilation of current electron-ion equilibration measurements at SNRs shocks has been presented in \cite{rakowski2005}, where an inverse relationship between the shock velocity and the level of temperatures equilibration has been claimed.
Namely, older SNRs, characterized by a lower value of the shock velocity, exhibit a $T_e/T_p$ ratio closer to, and in some cases consistent with 1.

It is important to stress that, even in the case of total non-equilibrium at the shock ($T_e/T_p = m_e/m_p$), electrons are unavoidably heated sufficiently far downstream of the shock through Coulomb collisions with hot thermal protons \citep{Spitzer}. This constitutes a sort of minimal heating scenario for electrons, which will be studied in details in this paper. The temperature of this electrons is measured directly from X-ray spectra \citep[e.g][]{Decourchelle2001,Vink2003}.

However, the electron heating through Coulomb interactions has been claimed to be too slow to explain the level of equilibration measured from optical lines observations in the vicinity of the shock for some SNRs \citep{Laming2000} and other mechanisms based on the generation of plasma waves are believed to proceed faster than Coulomb heating \citep[e.g.][]{Laming2001}.
Several of such collisionless mechanisms have been investigated, amongst others, by \cite{Cargill1988,Laming2001,Ghavamian2007,Rakowski2008}.

The efficient acceleration of cosmic rays at SNR shocks also plays a role, by reducing the total energy available for shock heating and consequently suppressing the proton temperature \citep{blasi2005,Ellison2007,Vink2008,Helder2008,Drury2009,Patnaude2009}.


The aim of this work is 
to present a simple model for electron Coulomb heating at SNR shocks, that could help to estimate the need for, and magnitude of, heating processes adjunctive to the Coulomb mechanism.  Our work is based on Chevalier's self-similar model which describes the evolution of SNRs in presence of effective cosmic ray acceleration at the SNR shock (\citealt{Chevalier1983}). We coupled Chevalier's model with the equations for electron heating by Coulomb exchange. The model lacks some important features; it does not include magnetic fields explicitly and the shock compression ratio is limited to a maximum value of 7 (since it does not include possibility for cosmic rays to escape from the shock), whereas in numerical simulations this ratio in some cases could reach much higher values (see for example \citealt{Ellison2005}).  However it does include a minimal set of the key physical processes and provides, we feel, a useful reference calculation for more complicated estimates and models.

The paper is organized as follows. In section \ref{sec:SNR dynamics} we briefly summarize the Chevalier model, introduce the equation for heating, and discuss projection effects in relating the 3D model to 2D observations. In section \ref{sec:Results} we summarize the results of the numerical solution of the equations of our model, and in section \ref{sec:Application} we apply the results of our model to the case of SN 1006. We conclude in Sec.~\ref{sec:Conclusions}.

\section{SNR dynamics}
\label{sec:SNR dynamics}
\subsection{Chevalier model}
A formal approach to the hydrodynamics of a SNR in adiabatic expansion phase including cosmic rays (CRs) acceleration was developed by \cite{Chevalier1983}.
\begin{figure}[ht!]
\includegraphics[width=.45\textwidth]{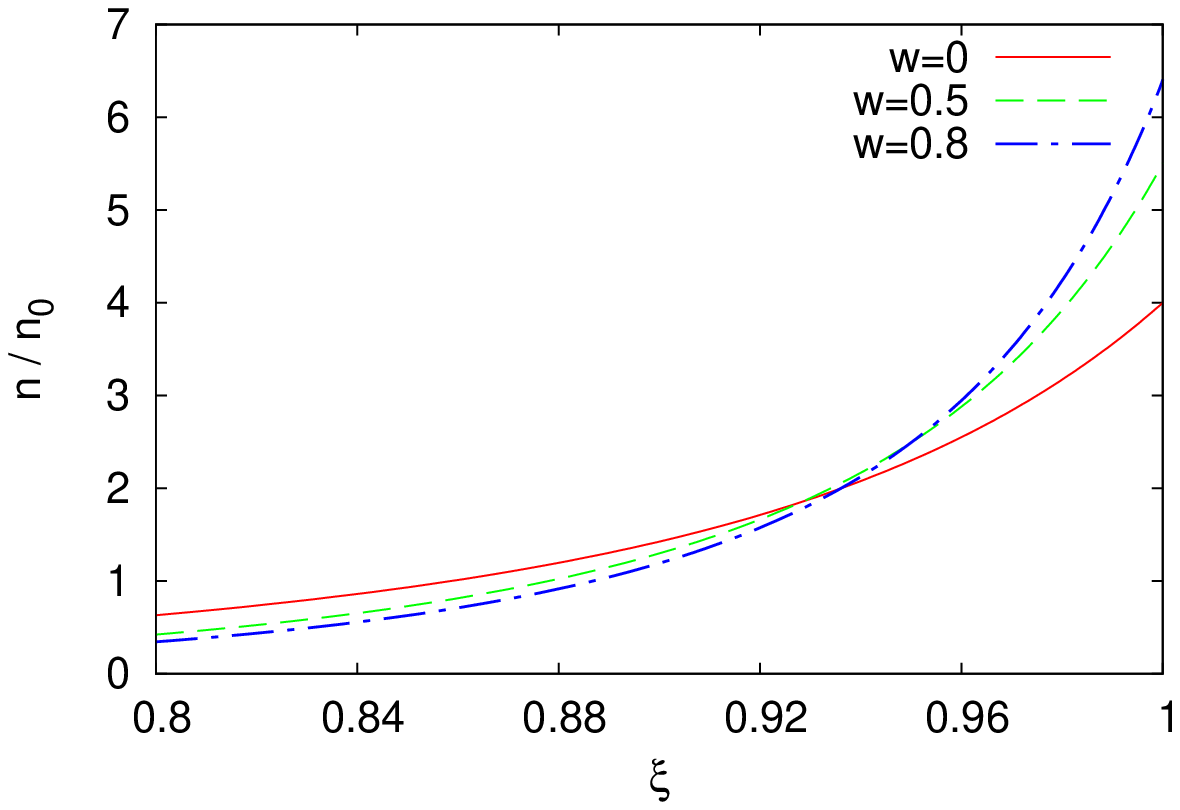}\\
\includegraphics[width=.45\textwidth]{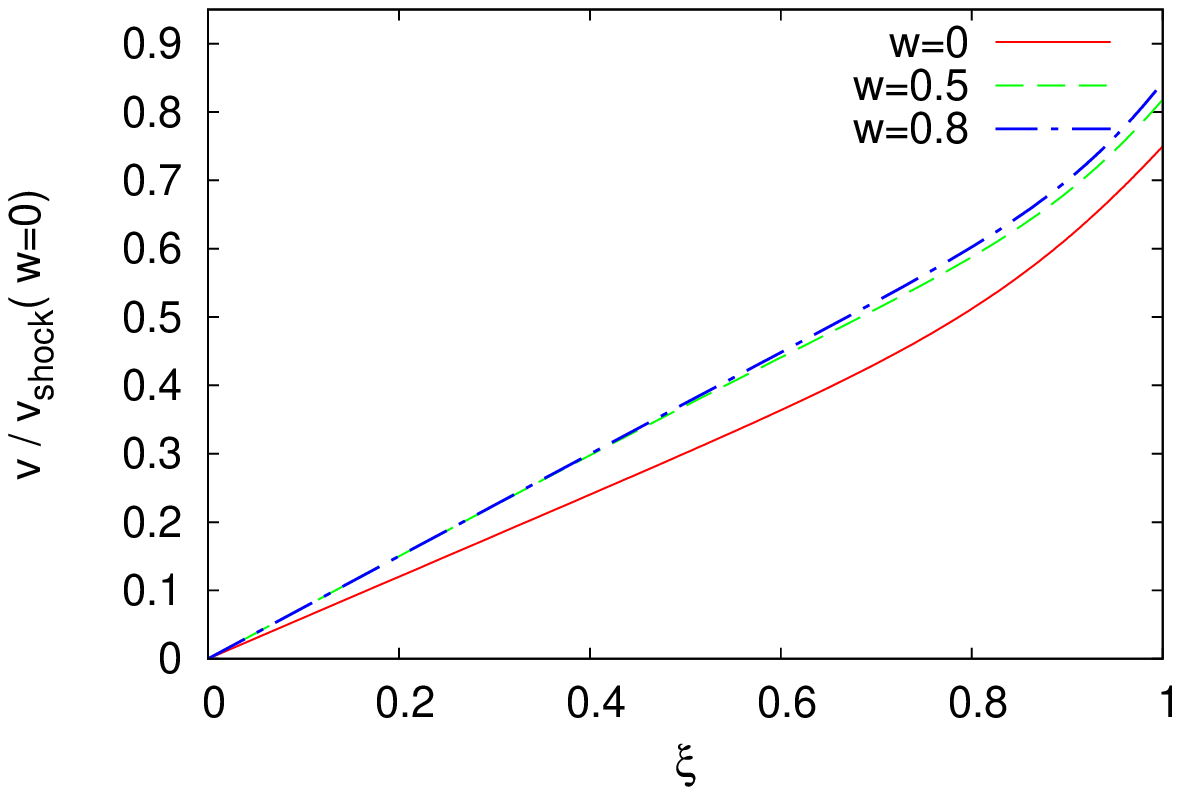}\\
\includegraphics[width=.45\textwidth]{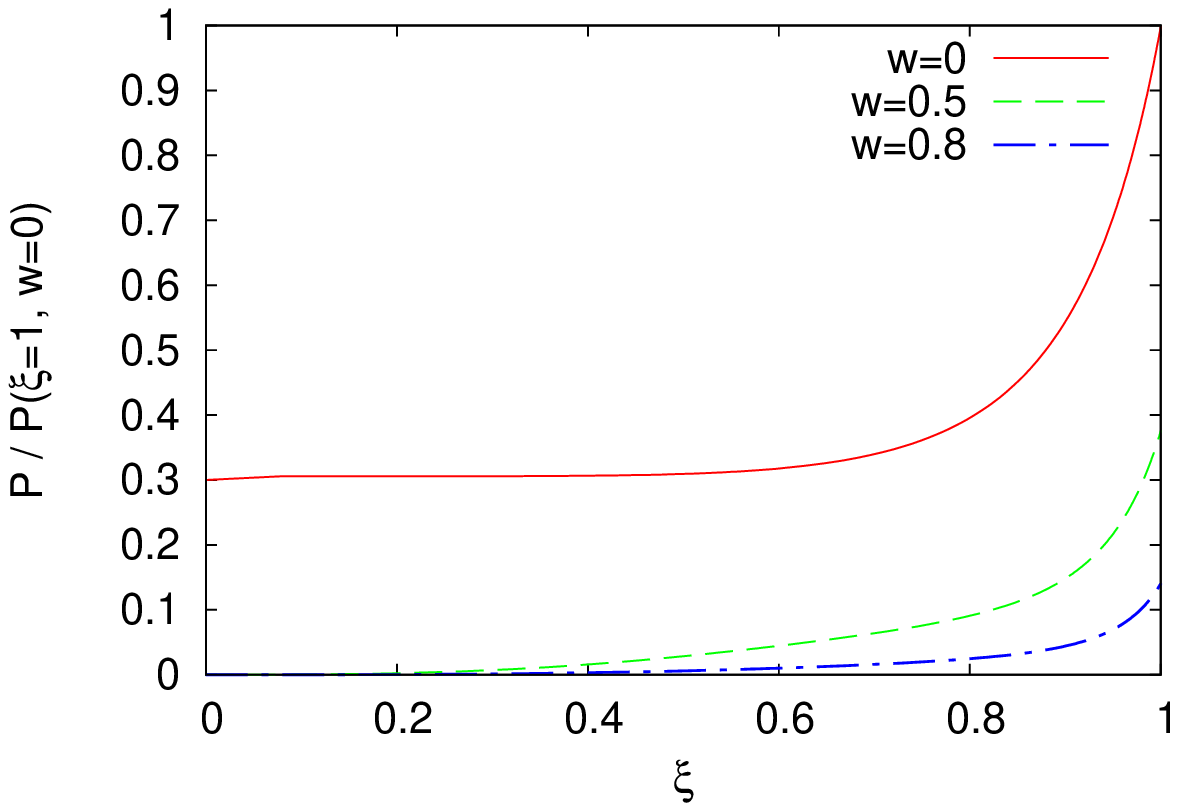}\\
\caption{Solutions of Chevalier model (see \citealt{Chevalier1983}) for density $n$, velocity $v$ and pressure $P_G$ inside the SNR. Different curves corresponds to different values of CR acceleration efficiency $w$. Here $\xi=r/r_{shock}$}
\label{fig:rho-v-p-chevalier}
\end{figure}

In his work Chevalier generalized the self-similar model of Sedov, which describes a point explosion in a uniform medium with negligible pressure.
Chevalier's model describes the interior of a spherically symmetric SNR by the two-fluid (relativistic cosmic rays and non-relativistic thermal gas) hydrodynamical equations without diffusion (including diffusion would in general break the self-similarity). In terms of the radial coordinate $0\le r\le r_{shock}(t)$ the equations are:
\begin{eqnarray}
\label{eqs-hydro-Chevalier}
&& \frac{\D \rho}{\D t}+v\frac{\D \rho}{\D r}+\rho\frac{\D v}{\D r}+\frac{2\rho v}{r} =0, \\ \nonumber
&& \frac{\D v}{\D t}+v\frac{\D v}{\D r}+\frac{1}{\rho}\frac{\D P_G}{\D r}+\frac{1}{\rho}\frac{\D P_C}{\D r}=0, \\ \nonumber
&& \frac{\D P_G}{\D t}+v\frac{\D P_G}{\D r}-\frac{5}{3}\frac{P_G}{\rho}\left(\frac{\D \rho}{\D t}+v\frac{\D \rho}{\D r}\right)=0, \\ \nonumber
&& \frac{\D P_C}{\D t}+v\frac{\D P_C}{\D r}-\frac{4}{3}\frac{P_C}{\rho}\left(\frac{\D \rho}{\D t}+v\frac{\D \rho}{\D r}\right)=0, \\ \nonumber
\label{eq-CR-modified-HD}
\end{eqnarray}
Here $\rho$ and $v$ stand for density and velocity of the gas, $P_C$ and $P_G$ pressures of the CRs and the thermal gas correspondingly. This system of equations describes mass and momentum conservation (first two equations) and entropy conservation for gas and cosmic rays (last two equations).

The boundary conditions come from applying generalized (i.e. cosmic rays included) jump conditions at the shock, which come from mass, momentum and energy conservation for particles that cross the shock, see \cite{Chevalier1983} for details.
To preserve self-similarity, the shock compression ratio, $s$, and the proportion of the total pressure going to cosmic rays $w$ (CR acceleration efficiency) must be taken as constants giving:
\begin{eqnarray}
\label{eqs-boundary-modified-2}
&& \rho(r_{shock}) = s \rho_0\\ \nonumber
&& v(r_{shock}) = \left(1-{1\over s}\right) \dot r_{shock}\\ \nonumber
&& P_C(r_{shock}) =  w \left(1-{1\over s}\right) \rho_0 \dot r_{shock}^2\\ \nonumber
&& P_G(r_{shock}) =  (1-w) \left(1-{1\over s}\right) \rho_0 \dot r_{shock}^2\\ \nonumber
\end{eqnarray}
In fact $s$ and $w$ should be taken from a nonlinear shock model, but at the level of this model it is enough to treat them as adjustable parameters.

In standard fashion the system of partial differential equations (\ref{eqs-hydro-Chevalier}) can now be reduced to a system of ordinary differential equations by transforming to the similarity variable
$\xi\propto rt^{-2/5}$, with the constant of proportionality chosen such such that $\xi=r/r_{shock}(t)$ (see \cite{Chevalier1983} for the constant's numerical value).

Introducing scaled structure functions $G$, $U$ and $Z$ we obtain
\begin{eqnarray}
\label{eqs-structure-functions}
&& \rho(r,t)=\rho_0G(\xi); \\ \nonumber
&& v(r,t)=\frac{2}{5}\frac{r}{t}U(\xi); \\ \nonumber
&& P(r,t)=\left(\frac{2}{5}\right)^2\left(\frac{r}{t}\right)^2\rho_0G(\xi)Z(\xi)\\ \nonumber
\end{eqnarray}
for the density, velocity, and pressure, respectively,
and for the compression ratio at the shock we get $s\equiv\frac{G(\xi=1+)}{G(\xi=1-)}=\frac{\gamma_w+1}{\gamma_w-1}$, where $\gamma_w=\frac{5+3w}{3(w+1)}$

In the following we will work in the coordinate system defined by the variables $\xi$ and $t$, so an element of fluid intersects the shock ($\xi=1$) and then moves with velocity $\dot{\xi}=\frac{2}{5}\xi(U(\xi)-1)$ towards the center ($\xi=0$) of the SNR.

The solution of these equations for different $w$ are shown in Fig.\ref{fig:rho-v-p-chevalier}, where $n=\rho/m_p$ is the particle number density.

\subsection{Electron heating}
\label{sec:electron heating}
We now consider Coulomb electron interactions with hot protons in the SNR model described in the previous subsection.
In the coordinate system that moves with the plasma, the equation that describes electron heating, can be found e.g. in \cite{Spitzer}. In the case of adiabatic expansion this equation reads:
 \begin{eqnarray}
 \label{eq-Spitzer-Te}
&& \frac{dT_e}{dt}=\frac{T_p-T_e}{t_{eq}}+\frac{\gamma-1}{n}T_e\frac{dn}{dt}; \\ \nonumber
&& t_{eq}=T_e^{3/2}\ln\Lambda/n_0; \\ \nonumber
 \end{eqnarray}
where $\ln\Lambda$ is the Coulomb logarithm.

The first term in Eq. \ref{eq-Spitzer-Te} corresponds to the heating of electrons due to Coulomb interactions with the protons of temperature $T_p$ and the second one corresponds to the cooling of the electron gas due to the adiabatic expansion of the SNR.
This equation has to be coupled with the equation that connects $T_p$ and $T_e$ with macroscopic parameters such as the gas pressure and density and with the equation that allows us to change coordinate system from the system moving with the gas to the system connected with the center of the explosion:
\begin{eqnarray}
\label{eq-model-def}
&& n(\xi)\left(T_e(\xi,t)+T_p(\xi,t)\right)=P_G(\xi,t); \\ \nonumber
&& \rho(\xi)=m_p n(\xi) \\ \nonumber
&& \dot{\xi}=\frac{2}{5}\xi(U(\xi)-1)\\ \nonumber
\end{eqnarray}
Together with the boundary condition $T_e(\xi=1,t)/T_p(\xi=1,t)=m_e/m_p$ this forms a complete set of equations for determining the electron temperature $T_e(\xi,t)$ inside a SNR.

\subsection{Projection effects}
\label{projection}

\begin{figure}[t!]
\resizebox{9.cm}{!}{\includegraphics[width=1.0\textwidth]{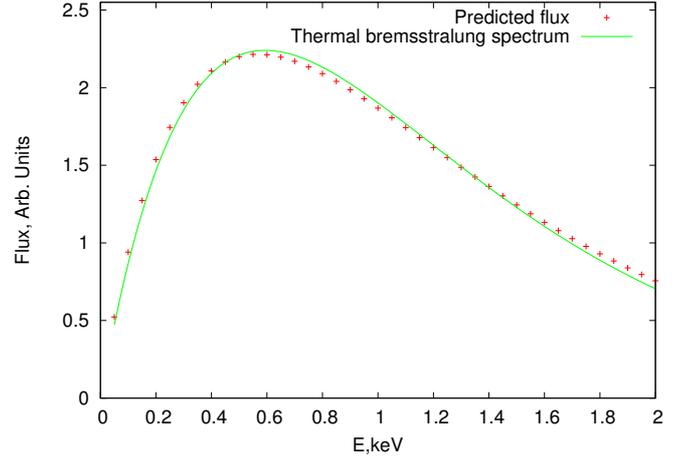}}
\caption{The points show the predicted spectra from area dS at  projected distance $\chi=0.7$ from the SNR center for a SNR age $t=500 \mbox{ yrs}$, explosion energy $E=10^{51}\mbox{ erg}$, gas density $n=0.1\mbox{ cm}^{-3}$, CR acceleration efficiency $w=0$. The solid line represents a single temperature TB spectra.}
\label{fig:spectra-chi=05}
\end{figure}

\begin{figure}[t!]
\resizebox{9.cm}{!}{\includegraphics[width=1.0\textwidth]{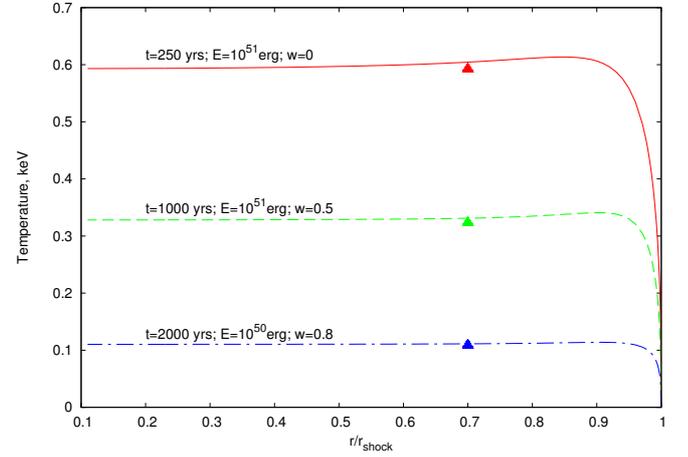}}
\caption{Electron temperature determined from the peak of the x-ray emission (see Fig.\ref{fig:spectra-chi=05}) as a function of the (projected) distance from the center of the SNR for different explosion energies $E$, ages $t$ and CR acceleration efficiencies $w$. The gas density is $n=0.1$ cm$^{-3}$ for all curves. Triangles indicate the temperature obtained by integrating the X-ray emission from the whole SNR.}
\label{fig:Te-chi}
\end{figure}

By solving Eq.~\ref{eq-Spitzer-Te} we derive the electron temperature $T_e(\xi,t)$ in the radial coordinate $\xi$. In order to compare this profile with observations we have to take into account projection effects. To do this we use the following formula for the thermal bremsstrahlung (TB) emission of the volume element $dV$ at distance $D$ (\citealt{Rybicki-Lightman}):
 \begin{eqnarray}
 \label{eq-TB-flux-Rybicki}
 \frac{dF_\nu}{1\mbox{keV}\ \mbox{s}^{-1}\ \mbox{cm}^{-2}}=720g_{ff}\left(\frac{D}{1\mbox{pc}}\right)^{-2} \left(\frac{n}{1\mbox{cm}^{-3}}\right)^2\times \\ \nonumber
 \times\left(\frac{T_e}{1\mbox{keV}}\right)^{-1/2}(\frac{E}{1\mbox{keV}})e^{-E/T_e}\left(\frac{dV}{1\mbox{pc}^3}\right)
 \end{eqnarray}
where $g_{ff}$ -- is the thermal Gaunt factor, $g_{ff}\approx 1$,  and the electron temperature $T_e$ here is a function of the distance $\xi$ to the SNR center.\\
The integration of this expression along the line of sight gives the flux from an area dS at a projected distance $\chi=r/r_{sh}$ from the SNR center.
The red points in Fig. \ref{fig:spectra-chi=05} represent the expected flux as a function of energy for $\chi=0.7$.
As it can be easily seen this spectrum is not exactly thermal, being the superposition of different thermal spectra characterized by different temperatures. However, a thermal bremsstrahlung spectrum gives a very good approximation to the predicted points. This is shown in Fig. \ref{fig:spectra-chi=05}, where the solid line represents the thermal bremsstrahlung emission for an effective temperature defined as the peak of the expected x-ray emission. The dependence of this effective temperature from $\chi$ is shown in Fig.~\ref{fig:Te-chi}.

This figure shows that, for a broad range of parameters such as explosion energy, SNR age and CR acceleration efficiency, such an effective electron temperature is almost constant in the inner ($\chi<0.9$) region of the SNR. So, it is reasonable to study the dependence of $T_e$ on other parameters not for all the values of $\chi$, but only for some fixed $\chi<0.9$. In the following we will adopt $\chi=0.7$.
Finally, the triangles in Fig.~\ref{fig:Te-chi} show the temperature one would observe by integrating the whole X-ray emission from the SNR. The discrepancy between the triangles and the curves is always below 5\%.

\begin{figure}
\includegraphics[width=0.47\textwidth]{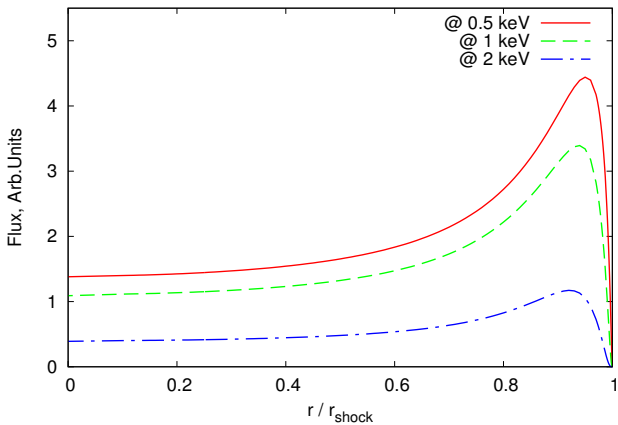}
\caption{Predicted total thermal flux profile for a SNR of age 1000~yr, explosion energy $10^{51}$~erg expanding in an ambient medium with density 0.1~cm$^{-3}$. The effects of CR acceleration has been neglected ($w = 0$). Solid red, dashed green and dot-dashed blue lines refer to the flux at energies equal to $0.5$, $1$ and $2$~keV, respectively.}
\label{fig:brightness profile}
\end{figure}

In Fig.~\ref{fig:brightness profile} we show the radial distribution of the thermal Bremsstrahlung X-ray flux for a SNR expanding in a medium with density $n = 0.1$~cm$^{-3}$. The SNR age is 1000~yr and the explosion energy is $E_{SN} = 10^{51}$~erg. We neglect the effect of CR acceleration at the SNR shock (i.e. $w = 0$).
The solid, dashed and dot--dashed lines refers to the emission profiles at 0.5, 1 and 2 keV, respectively.
The emission is roughly constant in the central region and exhibit a pronounced peak close to the position of the shock, indicating the presence of a shell-type morphology in thermal X-rays.

\begin{figure*}
{\includegraphics[width=0.5\textwidth]{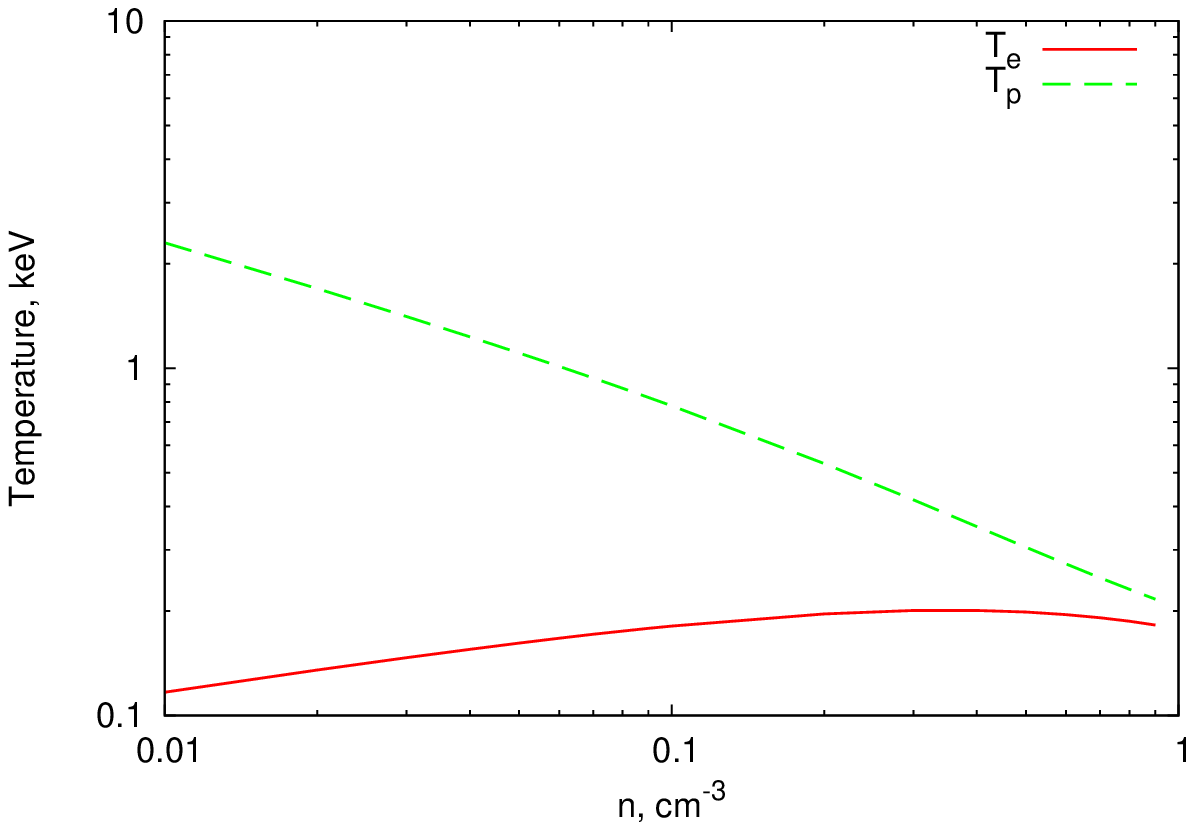}}
{\includegraphics[width=0.5\textwidth]{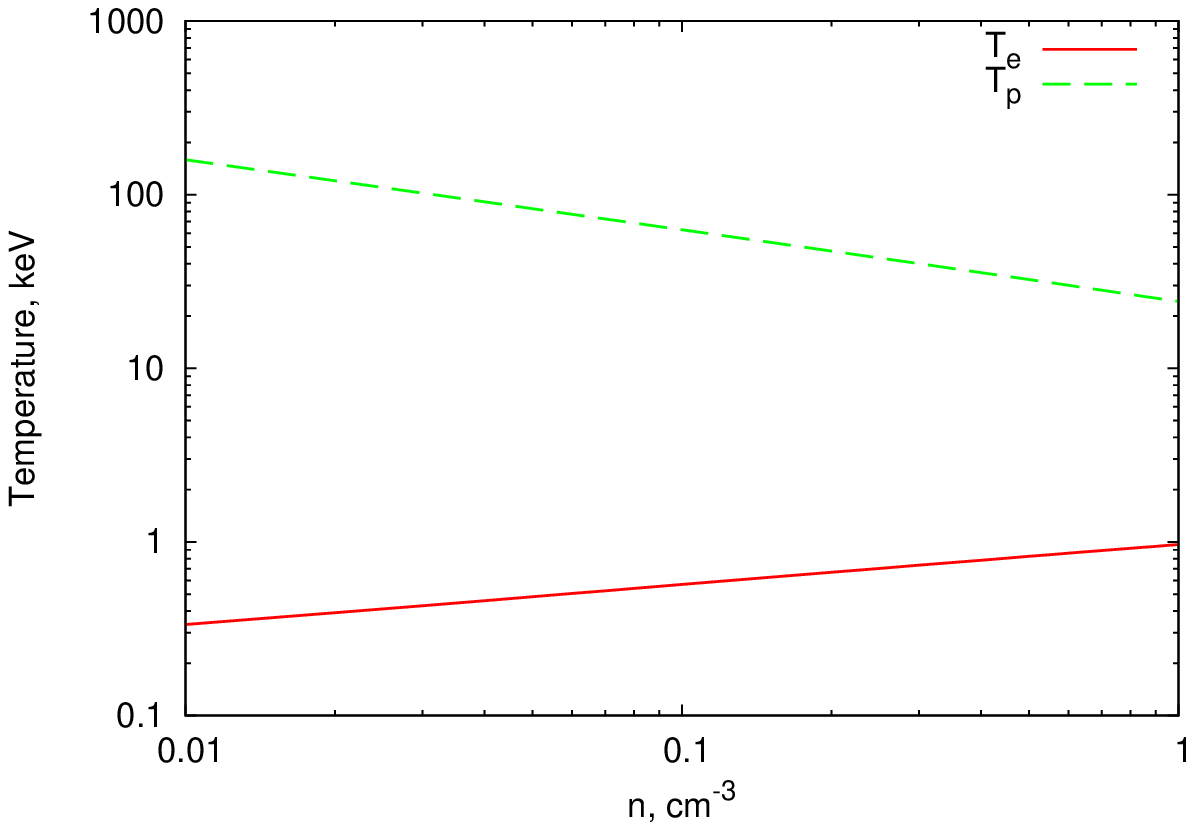}}
\caption{The left panel shows the electron and proton temperatures for a SNR of age $t=2000$ yr, CR acceleration efficiency $w=0.8$ and explosion energy $E_{SN}=10^{51}$erg.
The right panel shows analogous dependencies for a SNR for $t=500$yrs, $w=0$, $E_{SN}=10^{51}$erg.
In both cases projected temperatures are plotted, for a distance from the SNR centre equal to $r/r_{shock}=$0.7.}
\label{fig:te-tp-extreme}
\end{figure*}

\section{Results}
\label{sec:Results}

Here we solve numerically Eq.~\ref{eq-Spitzer-Te} and \ref{eq-model-def} to obtain the electron and proton temperature distributions inside a SNR. The dependence of  the projected temperatures $T_{e,p}(\chi = 0.7)$ (hereafter referred to as simply $T_{e,p}$) over parameters such as the upstream gas density $n$ and the supernova explosion energy $E_{SN}$ can be described as power laws for a broad range of parameter values ($n=0.01..1~\mbox{cm}^{-3}$, $t=125..2000$~yr, $E_{SN}=10^{50}..3\cdot10^{51}$~erg, $w=0..0.8$). Parameterizations for the electron temperature are given in tables \ref{table:Te(n)} and \ref{table:Te(E)} (see Appendix \ref{Appendix 1}). Here we briefly summarize the main characteristics of the solution.

Power law approximations provide a good fit to the exact solution except for the cases in which $T_e\sim T_p$. Fig. \ref{fig:te-tp-extreme} illustrates this situation: the electron and proton temperatures are shown as a  solid red and dashed green curve respectively, for the two cases $T_e \sim T_p$ (left panel) and $T_e\ll T_p$ (right panel).
The left panel shows the case of SNR of the age $t=2000$~yr with significant CR pressure ($w=0.8$). The explosion energy is $E_{SN}=10^{51}$erg. Under these assumptions for densities $n\gg0.1$~cm$^{-3}$, we get almost equal electron and proton temperatures. The dependence of these temperatures from the density deviates from a power law scaling.
The right panel of Fig. \ref{fig:te-tp-extreme} refers to a SNR with age $t=500$ yr, total explosion energy $E=10^{51}$erg and negligible acceleration of particles at the shock ($w=0$). In this case $T_e\ll T_p$ and as it is evident from the figure temperatures scale as a power laws of the density.

In general, results can be summarized as follows:
\begin{itemize}
  \item from $t=125$ yr to $t=2000$ yr and for $w<0.5$ the dependence of $T_e$ on gas density $n$ can be approximated with a single power law $T_e \propto n^b$ with index in the range $b = 0.15-0.23$ for n=0.01..1cm$^{-3}$. \\
For  $t\geq 1000$~yr and $w\geq 0.5$, the electron temperature $T_e$, in general, can not be approximated with a power law of the gas density (see Appendix \ref{Appendix 1} for details).

\item The dependence of $T_e$ on the total explosion energy $E_{SN}$ can be approximated as a power law $T_e \propto E_{SN}^b$ with index in the range $b = 0.16-0.31$. This is valid for the whole considered range of parameters (see Appendix \ref{Appendix 1}).
\end{itemize}

It has been argued that some additional mechanism of electron heating based on the generation of plasma waves must operate in SNRs \citep{Cargill1988,Laming2001,Ghavamian2007,Rakowski2008}.

Apriori, two possibilities for this mechanisms could be considered. They might operate only in the vicinity of the shock \citep{Cargill1988,Laming2001,Ghavamian2007,Rakowski2008}, either, similar to Coulomb heating mechanism, operate all way downstream. Some physical justification for the second type mechanisms could be found in \cite{Laming2004,Laming2007} and \cite{Sharma2007}, were analogous processes for the solar wind and hot accretion flows were considered.

Formally speaking, mechanisms operating very close to the shock, implies changing of boundary conditions for our equations. We will consider them in details for SN~1006 in the section \ref{sec:Application}. Results for mechanism that operate all way downstream are briefly summarized below.

If we assume for simplicity that this mechanism has the same temperature dependence as the Coulomb one, but is $k$ times faster, then the heating of electrons can be described by an equation similar to Eq.~\ref{eq-Spitzer-Te}:
$$\frac{dT_e}{dt}=k\frac{T_p-T_e}{t_{eq}}+\frac{\gamma-1}{n}T_e\frac{dn}{dt}.$$
By solving this equation we found that the electron temperature scales with $k$ as $T_e(k)\sim k^{0.8-0.9}$.

 A comparison between the predictions of our model, which considers only Coulomb heating, and the observed SNR electron temperatures can serve as a tool to estimate whether such additional heating mechanisms are needed at all, and, if so, at which level.

\section{Application of the model to the supernova remnant SN 1006}
\label{sec:Application}

In order to compare the predictions of our model with observations we consider here the well known SNR SN 1006.
The choice of this object is justified by the fact that it is a very well studied SNR, for which a relatively accurate determination of parameters such as distance, shock velocity and age, is available.

SN 1006 is an historical SNR with age $\approx 1000$~yr and believed to be in the Sedov phase of its evolution.
After reviewing the preexisting literature, \cite{Dubner2002} concluded that the distance to SN 1006 lies in the range 1.4--2 kpc.
More recently, a slightly larger distance of $2.18$~kpc was inferred by \cite{Winkler2003}, who compared the measured optical proper motion with the expansion velocity of $2890 \pm 100$~km/s determined from H$\alpha$ lines observations. This velocity is in a good agreement with analogous velocity measurements by \cite{Smith1991} which give $v=2200-3500$~km/s.
At this distance the SNR apparent size 15' (\citealt{Rothenflug2004}) translates into a shock radius equal to $r_{shock}=9.6$~pc.
A value for the ambient gas density equal to $n=0.15-0.25$~cm$^{-3}$ (in the north-western region) has been obtained by \cite{Acero2007} from XMM X-ray observations. A slightly higher values for the density $n=0.25-0.4$~cm$^{-3}$ was obtained from H$\alpha$ lines observations with HST (\citealt{Raymond2007}).

By assuming that the evolution of SN~1006 can be described by the Chevalier self--similar model described in Sec.~\ref{sec:SNR dynamics}, it is possible to derive a supernova explosion energy equal to $E_{SN} \approx 2 \times 10^{51}$~erg.

\begin{figure}
{\includegraphics[width=0.5\textwidth]{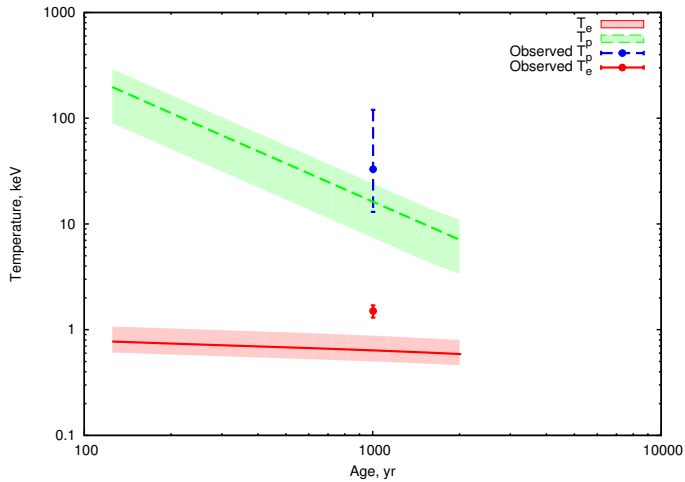}}
\caption{Solid and dashed lines show the electron and proton temperature for SN 1006 as a function of the SNR age. Shaded regions represent the model uncertainty (see text for details). Data points refer to the observed values of $T_e$ and $T_p$ (dots with error bars) adopted from \cite{Vink2003}, see text for the details.}
\label{fig:te-tp-sn1006}
\end{figure}

Results from our model for SN~1006 are shown in Fig.~\ref{fig:te-tp-sn1006}, where the electron and proton temperatures behind the SNR shock wave are plotted as a function of the remnant age.
For the actual age of the remnant, $t \approx 1000$~yr, ambient gas density $n=0.2$~cm$^{-3}$, explosion energy $E=2\cdot 10^{51}$~erg and $w=0.1$, our model gives a value of the proton temperature equal to $T_p \sim 17$~keV, and an electron temperature equal to $T_e = 0.6$~keV.
The observational values are adopted\footnote{Note, that in \cite{Vink2003} not the proton, but oxygen temperature is measured directly. We adopt here $T_p=m_p/m_O T_O=1/16T_O\approx 33$~keV} from \cite{Vink2003} for the proton and electron temperatures are also indicated in Fig.~\ref{fig:te-tp-sn1006}, together with the 
error bars ($3 \sigma$ for the proton temperature and 90\% confidence level for the electron temperature). A bit lower proton temperature $T_p=1.8\cdot 10^8K\approx 16keV$ was obtained by \cite{Korreck2004}.
It is clear from Fig.~\ref{fig:te-tp-sn1006} that our model predicts a proton temperature which is in agreement with observations, while the prediction for the electron temperature falls a factor of 2--3 below the observational value of $T_e\approx 1.5\pm 0.2$~keV \citep{Vink2003}, (XMM).
A similar value for the electron temperature of $T_e = 1.5 - 1.7$~keV has been independently measured by \cite{Acero2007} with XMM.
Nevertheless there is some doubt in the electron temperature measurements. Lower (0.6-0.7keV) values were obtained by \cite{Long2003} with CHANDRA for NW-1 region, while other NW regions appear to be ejecta dominated. Since XMM has worse spatial resolution than CHANDRA, ejecta dominated regions can contribute significantly to its spectra. Moreover, in \cite{Acero2007} it is shown that including synchrotron emission component can reduce observed electron temperature.

If density and velocity vary in the range $n=0.15-0.5$~cm$^{-3}$, $v=2400-3500$~km/s, respectively, we estimated the corresponding range of electron temperatures predictions from the model to be $T_e=0.5-0.9$~keV. This values are in the agreement with ones from the model in \cite{Vink2003}, see their Fig. 2 .
The shaded regions in Fig.~\ref{fig:te-tp-sn1006} represent the predictions from our model by taking into account the above mentioned uncertainties on SNR parameters.

Thus, the predicted electron temperature agrees well with fits to NW-1 region in \cite{Long2003}, but smaller, than the temperatures obtained by \cite{Vink2003,Acero2007}. Part of the discrepancy between the measured and expected electron temperature might be explained by the uncertainties in the determination of SNR parameters such as the external gas density or the shock velocity.

\begin{figure}
{\includegraphics[width=0.5\textwidth]{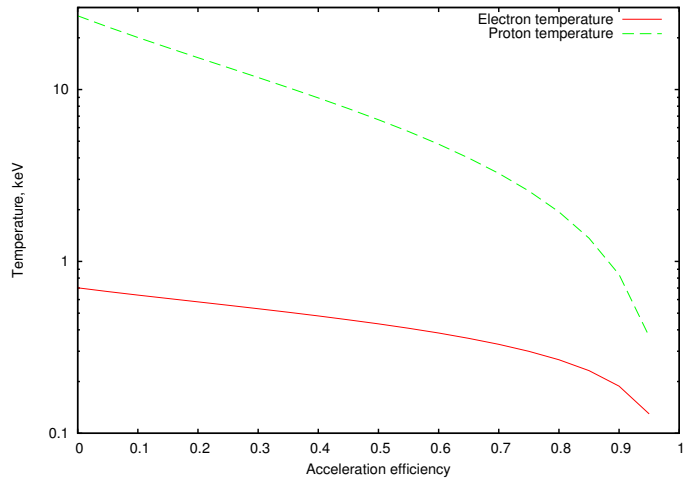}}
\caption{Solid and dashed lines show the electron and proton temperature for SN 1006 as a function of acceleration efficiency $w$.}
\label{fig:te-tp-w-sn1006}
\end{figure}

Dependence of the electron and proton temperature on acceleration efficiency for SN~1006 is shown in Fig.~\ref{fig:te-tp-w-sn1006}. Both temperatures quickly decreasing with increasing $w$. The result shown in Fig.~\ref{fig:te-tp-sn1006} corresponds to $w=0.05$.

Recent detection of SN~1006 NE and SW regions at TeV gamma rays by H.E.S.S. \citep{HESSsn1006} implies that CRs protons and/or electrons are accelerated at the SNR shock. The density $n_0\approx 0.085$~cm$^{-3}$ \citep{Katsuda2009} in this regions is lower than in NW and thus they can be in early Sedov or late free-expansion phase. The measurements of expansion index by \cite{Katsuda2009} give $m=0.54\pm 0.05$, i.e. considerably less than if it were in a free expansion ($m=1$), but higher than for an adiabatically expanding SNR ($m=0.4$).
For an estimation of the electron temperature we consider this SNR to be in Sedov phase.
Under the assumption that the observed TeV emission has an hadronic origin it is possible to estimate the CR acceleration efficiency to be roughly equal to $w \approx 0.2$ \citep{HESSsn1006}. For efficiency range $w=0.1..0.2$ our model gives electron temperature $T_e\approx 0.47-0.52$~keV.

The expected electron temperature can be further increased if additional heating mechanisms for electrons operate at SNR shocks. It has been suggested that these mechanisms might be related to the presence of plasma waves upstream of the shock that can effectively heat electrons \citep{Cargill1988,Laming2001,Ghavamian2007,Rakowski2008}. The fact that such heating is expected to happen upstream of the shock is suggested by optical observations of H$\alpha$ lines from a number of SNR shocks \citep[see][and references therein]{Ghavamian2007}.
These observations show that the electron temperature might be already significantly heated above the expected value of $T_e = (m_e/m_p) ~ T_p$ immediately after the shock, since the width of H$\alpha$ filaments is of the order of $\approx 10^{14}$~cm.
It is important to stress that, on the other hand, the temperatures measured from X-ray observations of SNRs refers to a quite broad region downstream of the shock (e.g. $\lesssim$10\%$r_{sh}$ for SN 1006).

In order to estimate the effect of additional heating happening upstream of the shock, we repeated our calculations for SN~1006 by assuming that the ratio of the proton to the electron temperature immediately after the shock is equal to $T_e/T_p = 1/15$, instead of the canonical value $T_e/T_p = m_e/m_p \approx 1/1800$ assumed in Fig.~\ref{fig:te-tp-sn1006}.
Such value is consistent with the range of values derived from H$\alpha$ measurements $T_e/T_p \lesssim 0.07$ \citep{Ghavamian2002,Ghavamian2007}. Thus, this is equivalent to assume that the additional heating increases the electron temperature at the shock by a factor of $\approx$~100. For such a choice of the boundary condition our model predicts a value for the effective electron temperature (the one that is measured from X-ray observations) equal to 1.4 keV. This value is higher than the one obtained by the minimal heating model shown in Fig.~\ref{fig:te-tp-sn1006} and in agreement with observations \citep{Vink2003}.
It is important to stress that an increase of about two orders of magnitude of the heating of electrons at the shock (which leads to $T_e/T_p = 1/15$ immediately downstream of the shock), is reflected by a moderate increase (a factor of $\approx$~2 only) of the predicted effective temperature which can be measured from X-ray observations.
This is due to the fact that the electron temperature well downstream of the shock is mainly determined by an equilibrium between Coulomb heating and adiabatic losses and depends only mildly on the boundary condition for $T_e/T_p$ at the shock.



\section{Discussion and conclusions}
\label{sec:Conclusions}

In this paper, we studied the evolution in time of the proton and electron temperatures behind the shock waves of SNRs in the Sedov phase of their evolution. To describe the evolution of the SNR we adopted the model described by \cite{Chevalier1983}, which generalizes the Sedov solution by allowing the acceleration of CRs at the SNR shock. Immediately after being shocked, the proton gas is heated to the temperature given by the modified (by the inclusion of CR acceleration) Rankine-Hugoniot relations, while the electron temperature was assumed to be a factor of $m_p/m_e$ lower.
Downstream of the shock electrons can be further heated due to Coulomb collisions with the hot protons.

Since Coulomb heating is unavoidable and is the only heating process included in our calculations, our results has to be intended as estimates of the minimal electron temperatures expected in SNRs.

We applied the model to the case of SN 1006. The model predicts a value of the electron temperature equal to $T_e=0.6$~keV, roughly a factor of 2 smaller than the value measured from X-ray data, which is $T_e=1.5$~keV. The discrepancy may be reduced by taking into account the uncertainties in the determination of the SNR parameters such as the shock speed or the external gas density.

In fact, other heating process based for example on the generation of plasma waves upstream of the shock may play a role \citep[see e.g.][]{Rakowski2008}. Such additional collisionless processes have been sometimes invoked to explain the observed electron temperatures, since Coulomb heating alone seemed to be too slow in certain cases \citep{Laming2000}.
In this respect, our model can serve as a useful tool to estimate the order of magnitude of such extra heating, when that is required.
For the considered case of SN~1006, we showed that the discrepancy of a factor of $\approx$~2 between the expected electron temperature and the one measured from X-ray observations can be indeed compensated if an additional heating mechanism operates upstream of the shock. This mechanism has to be very effective and able to increase the electron temperature immediately downstream of the shock (e.g. before Coulomb heating starts to operate) of about two orders of magnitudes above the canonical value of $T_e = (m_e/m_p) ~ T_p$.

\section*{Acknowledgements}
We thank the referee for a prompt and constructive report that led to an improved paper. The work of D.M. was supported by grant 07/RFP/PHYF761 from Science Foundation Ireland (SFI) under its Research Frontiers Programme.
The authors wish to acknowledge the SFI/HEA Irish Centre for High-End Computing (ICHEC) for the provision of computational facilities and support.

\clearpage
\begin{appendix}
\section{Power law fits for the electron temperature}
\label{Appendix 1}
In this Appendix we present the coefficients for the power law fits of the effective electron temperature $T_e$ as defined in Sec.~\ref{projection}. For fixed values of the SNR age $t$ and CR acceleration efficiency $w$, the electron temperature $T_e$ is fitted as a power law function of the ambient gas density $n$ and explosion energy $E_{SN}$. Fits parameters are given for values of $w$ in the range $0..0.8$ and for vlaues of the SNR age in the range $125..2000$ yrs.

Table \ref{table:Te(n)} gives the values of the parameters for the power law fits:
$$T_e(n)/1\mbox{keV}=a\cdot(n/1\mbox{cm}^{-3})^{b}.$$
All fits have an accuracy better than $\approx$10\% for values of  the ambient density in the range $0.01..1$~cm$^{-3}$. For some sets of parameters a power law fit could not give a satisfactory approximation (better than 10\%) for $T_e$. For these cases the electron temperature $T_e(n)$ is plotted in Fig.~A.1 and Fig.~A.2. Each curve in the figures represents $T_e(n)$ for different values of explosion energy, see figures captions for the details.

Table \ref{table:Te(E)} gives the values of the parameters for the power law fits:
$$T_e(E)/1\mbox{keV}=a\cdot(E/10^{51}\mbox{erg})^b.$$
The accuracy of fits is better than $\approx 10$~\% when the value of the ambient density is in the range $n=0.01..1$~cm$^{-3}$ and the CR acceleration efficiency is int he range $w=0..0.8$.

\begin{table*}[h!]
\begin{tabular}{|c|c|c|c|c|c|c|c|c|c|c|c|c|}
  \hline
  E$\rightarrow$ & \multicolumn{2}{|c|}{$10^{50}$} &  \multicolumn{2}{c|}{$3\cdot 10^{50}$}&  \multicolumn{2}{c|}{$5\cdot 10^{50}$}&  \multicolumn{2}{c|}{$7\cdot 10^{50}$}&  \multicolumn{2}{c|}{$10^{51}$}&  \multicolumn{2}{c|}{$3\cdot 10^{51}$}\\
  \hline
  \multicolumn{13}{|c|}{w=0}\\
  \hline
  t$\downarrow$&a&b&a&b&a&b&a&b&a&b&a&b\\
  \hline
  125& 0.75 & 0.217 & 0.89 & 0.198 & 0.96 & 0.194 & 1.01 &0.187 & 1.06 &0.178&1.25&0.145\\
  \hline
  250& 0.70 & 0.230 & 0.84 & 0.228  & 0.92 &0.227 & 0.97 & 0.225 & 1.03 & 0.224&1.23&0.216 \\
  \hline
  500& 0.65 & 0.231 & 0.78 & 0.234 & 0.86 &0.230 &0.91 &0.232 & 0.96 & 0.230&1.16&0.231 \\
  \hline
  1000& 0.59 & 0.215 & 0.72 & 0.221 & 0.79 & 0.226 & 0.83 & 0.224 & 0.89 & 0.225 & 1.07 &0.227 \\
  \hline
  2000& 0.50 &0.187 & 0.63 &0.199 & 0.69 &0.204 & 0.74 & 0.207 & 0.79 & 0.209&0.97 &0.216 \\
  \hline
  \multicolumn{13}{|c|}{w=0.3}\\
  \hline
  125& 0.57 &0.229&0.68& 0.225&0.74&0.222&0.79&0.220&0.83&0.217&0.99&0.203 \\
  \hline
  250&0.44&0.226&0.53&0.228&0.57&0.229&0.61&0.230&0.64&0.230&0.94&0.228 \\
  \hline
  500&0.49&0.220&0.59&0.224&0.65&0.226&0.68&0.227&0.73&0.227&0.88&0.229 \\
  \hline
  1000&0.42&0.198&0.52&0.208&0.58&0.211&0.62&0.213&0.66&0.215&0.80&0.221 \\
  \hline
  2000&0.33&0.148&0.43&0.170&0.48&0.178&0.52&0.183&0.56&0.187&0.70&0.199\\
  \hline
  \multicolumn{13}{|c|}{w=0.5}\\
  \hline
  125&0.46&0.231&0.56&0.230&0.61&0.228&0.65&0.228&0.69&0.226&0.82&0.221\\
  \hline
  250&0.44&0.226&0.53&0.228&0.57&0.229&0.61&0.230&0.64&0.230&0.77&0.231\\
  \hline
  500&0.39&0.212&0.48&0.218&0.52&0.220&0.56&0.222&0.59&0.223&0.72&0.226\\
  \hline
  1000&0.32&0.180&0.41&0.193&0.45&0.199&0.49&0.202&0.52&0.205&0.64&0.213\\
  \hline
  2000&--&--&--&--&--&--&--&--&--&--&0.55&0.178\\
  \hline
  \multicolumn{13}{|c|}{w=0.8}\\
  \hline
  125&0.30&0.226&0.36&0.228&0.39&0.229&0.41&0.230&0.44&0.230&0.53&0.231\\
  \hline
  250&0.27&0.213&0.33&0.219&0.36&0.221&0.38&0.222&0.41&0.223&0.49&0.227\\
  \hline
  500&0.23&0.183&0.28&0.196&0.31&0.201&0.34&0.204&0.36&0.207&0.44&0.214\\
  \hline
  1000&--&--&--&--&--&--&--&--&--&--&0.38&0181\\
  \hline
  2000&--&--&--&--&--&--&--&--&--&--&--&--\\
  \hline
  \end{tabular}
  \caption{Fit for the electron temperature $T_e$ for SNR inner regions (i.e. at projected distance $r/r_{shock}\lesssim0.8$) as function of outer density $T_e(n)/1\mbox{keV}=a\cdot(n/1\mbox{cm}^{-3})^{b}$, for different explosion energies, if applicable. See Fig.A.1, Fig.A.2 otherwise. Uncertainties of all fits are less than $\approx$10\% }
  \label{table:Te(n)}
\end{table*}
\begin{figure*}[b!]
\begin{tabular}[c]{cccc}
\TwoLines{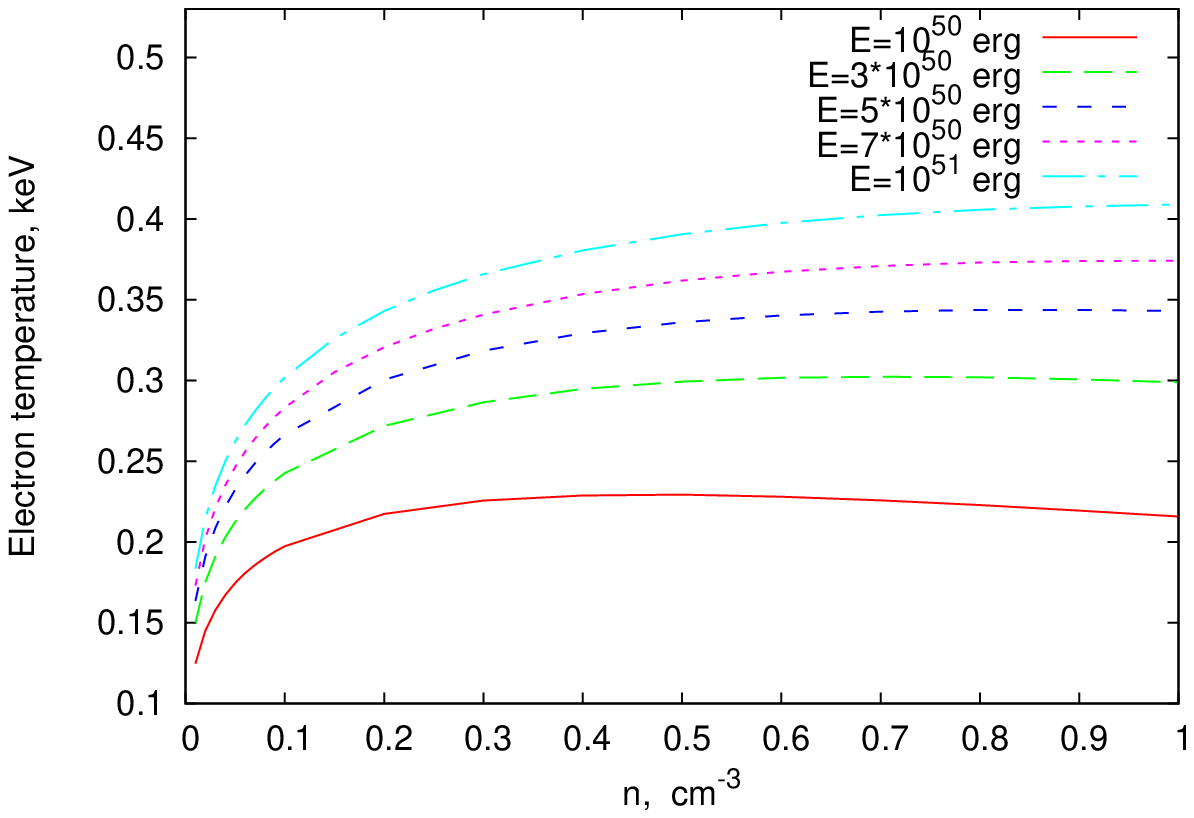}{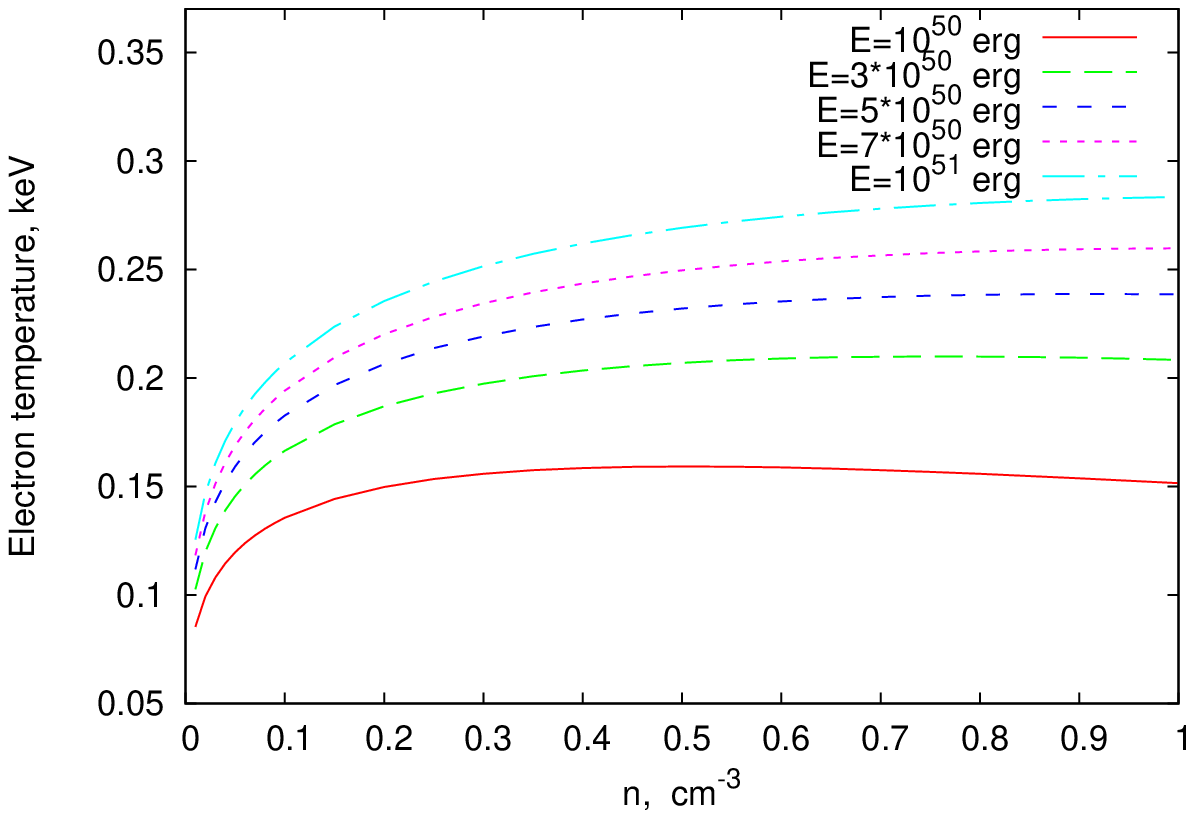}
\end{tabular}
\caption{Electron temperature $T_e$ for SNR inner regions (i.e. at projected distance $r/r_{shock}\lesssim0.8$) for parameters for which powerlaw fit is not applicable (see table \ref{table:Te(n)}).
 Left panel: $T_e(n)$ dependence for t=2000, w=0.5. Each curve represent different explosion energy.
 Right panel: $T_e(n)$ dependence for t=1000, w=0.8. Each curve represent different explosion energy.}
\end{figure*}
\begin{table*}[]
  \begin{tabular}{|c|c|c|c|c|c|c|c|c|c|c|}
  \hline
  n$\rightarrow$ & \multicolumn{2}{|c|}{$0.01$} &  \multicolumn{2}{c|}{$0.05$}&  \multicolumn{2}{c|}{$0.1$}&  \multicolumn{2}{c|}{$0.5$}&  \multicolumn{2}{c|}{$1$}\\
  \hline
  \multicolumn{11}{|c|}{w=0}\\
  \hline
  t$\downarrow$&a&b&a&b&a&b&a&b&a&b\\
  \hline
125  &  0.539  &  0.269  &  0.576  &  0.181  &  0.651  &  0.170  &  0.936  &  0.165  &  1.097  &  0.165\\
\hline
250  &  0.375  &  0.194  &  0.513  &  0.166  &  0.599  &  0.165  &  0.877  &  0.165  &  1.026  &  0.166\\
\hline
500  &  0.339  &  0.168  &  0.481  &  0.164  &  0.570  &  0.165  &  0.824  &  0.168  &  0.964  &  0.171\\
\hline
1000  &  0.307  &  0.165  &  0.453  &  0.165  &  0.529  &  0.167  &  0.761  &  0.174  &  0.893  &  0.180\\
\hline
2000  &  0.292  &  0.165  &  0.427  &  0.168  &  0.504  &  0.171  &  0.692  &  0.187  &  0.782  &  0.203\\
  \hline
  \multicolumn{11}{|c|}{w=0.3}\\
  \hline
125  &  0.327  &  0.213  &  0.428  &  0.169  &  0.497  &  0.165  &  0.716  &  0.165  &  0.835  &  0.166\\
\hline
250  &  0.269  &  0.174  &  0.395  &  0.165  &  0.453  &  0.165  &  0.676  &  0.167  &  0.782  &  0.169\\
\hline
500  &  0.249  &  0.165  &  0.366  &  0.165  &  0.440  &  0.166  &  0.624  &  0.172  &  0.720  &  0.176\\
\hline
1000  &  0.234  &  0.165  &  0.353  &  0.167  &  0.398  &  0.169  &  0.578  &  0.182  &  0.657  &  0.194\\
\hline
2000  &  0.220  &  0.166  &  0.327  &  0.172  &  0.376  &  0.177  &  0.499  &  0.207  &  0.552  &  0.237\\
  \hline
  \multicolumn{11}{|c|}{w=0.5}\\
  \hline
125  &  0.247  &  0.189  &  0.337  &  0.166  &  0.409  &  0.165  &  0.586  &  0.166  &  0.690  &  0.167\\
\hline
250  &  0.221  &  0.168  &  0.322  &  0.165  &  0.380  &  0.165  &  0.546  &  0.169  &  0.637  &  0.171\\
\hline
500  &  0.205  &  0.165  &  0.293  &  0.166  &  0.353  &  0.167  &  0.509  &  0.175  &  0.594  &  0.183\\
\hline
1000  &  0.190  &  0.165  &  0.280  &  0.169  &  0.328  &  0.173  &  0.466  &  0.191  &  0.518  &  0.209\\
\hline
2000  &  0.177  &  0.168  &  9.782  &  1.760  &  0.306  &  0.184  &  0.391  &  0.230  &  0.413  &  0.274 \\
  \hline
  \multicolumn{11}{|c|}{w=0.8}\\
  \hline
125  &  0.147  &  0.168  &  0.219  &  0.165  &  0.263  &  0.165  &  0.368  &  0.168  &  0.445  &  0.171\\
\hline
250  &  0.143  &  0.165  &  0.205  &  0.166  &  0.250  &  0.167  &  0.344  &  0.175  &  0.410  &  0.181\\
\hline
500  &  0.135  &  0.165  &  0.192  &  0.169  &  0.222  &  0.170  &  31.286  &  0.190  &  0.354  &  0.206\\
\hline
1000  &  0.125  &  0.168  &  0.180  &  0.176  &  0.213  &  0.183  &  0.269  &  0.226  &  0.279  &  0.269\\
\hline
2000  &  0.116  &  0.173  &  0.156  &  0.192  &  0.179  &  0.211  &  0.205  &  0.270  &  0.185  &  0.314\\
  \hline
  \end{tabular}
  \caption{Fit for the electron temperature $T_e$ for SNR inner regions (i.e. at projected distance $r/r_{shock}\lesssim0.8$) as function of explosion energy $T_e(E)/1\mbox{keV}=a\cdot(E/10^{51}\mbox{erg})^b$ for different outer gas densities. Uncertainties of all fits are less than 10\%}
  \label{table:Te(E)}
\end{table*}

\begin{figure*}[p!]
{\includegraphics[width=0.47\textwidth]{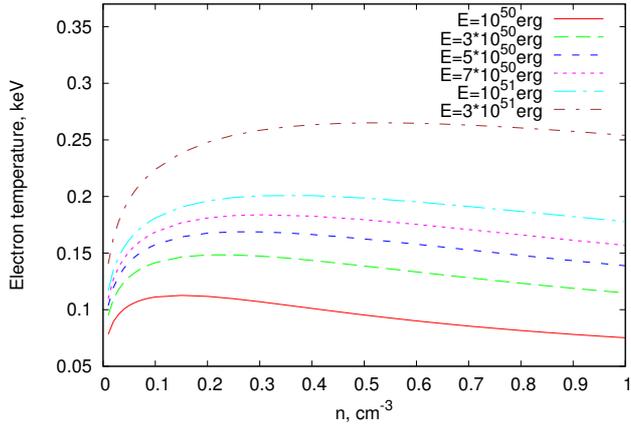}}
\caption{Same as Fig.A.1 for t=2000, w=0.8}
\end{figure*}
\end{appendix}
\clearpage
\bibliography{astro}

\begin{thebibliography}{34}
\expandafter\ifx\csname natexlab\endcsname\relax\def\natexlab#1{#1}\fi

\bibitem[{{Acero} {et~al.}(2007){Acero}, {Ballet}, \&
  {Decourchelle}}]{Acero2007}
{Acero}, F., {Ballet}, J., \& {Decourchelle}, A. 2007, \aap, 475, 883

\bibitem[{{Aharonian} {et~al.}(2010){Aharonian}, {Akhperjanian}, {Anton},
  {Barres de Almeida}, {Bazer-Bachi}, {Becherini}, {Behera}, {Beilicke},
  {Bernl{\"o}hr}, {Bochow}, {Boisson}, {Bolmont}, {Borrel}, {Brucker}, {Brun},
  {Brun}, {B{\"u}hler}, {Bulik}, {B{\"u}sching}, {Boutelier}, {Chadwick},
  {Charbonnier}, {Chaves}, {Cheesebrough}, {Conrad}, {Chounet}, {Clapson},
  {Coignet}, {Dalton}, {Daniel}, {Davids}, {Degrange}, {Deil}, {Dickinson},
  {Djannati-Ata{\"i}}, {Domainko}, {O'C.~Drury}, {Dubois}, {Dubus}, {Dyks},
  {Dyrda}, {Egberts}, {Eger}, {Espigat}, {Fallon}, {Farnier}, {Fegan},
  {Feinstein}, {Fiasson}, {F{\"o}rster}, {Fontaine}, {F{\"u}{\ss}ling},
  {Gabici}, {Gallant}, {G{\'e}rard}, {Gerbig}, {Giebels}, {Glicenstein},
  {Gl{\"u}ck}, {Goret}, {G{\"o}ring}, {Hauser}, {Hauser}, {Heinz},
  {Heinzelmann}, {Henri}, {Hermann}, {Hinton}, {Hoffmann}, {Hofmann},
  {Hofverberg}, {Holleran}, {Hoppe}, {Horns}, {Jacholkowska}, {de Jager},
  {Jahn}, {Jung}, {Katarzy'nski}, {Katz}, {Kaufmann}, {Kerschhaggl},
  {Khangulyan}, {Kh{\'e}lifi}, {Keogh}, {Klochkov}, {Klu'zniak}, {Kneiske},
  {Komin}, {Kosack}, {Kossakowski}, {Lamanna}, {Lemoine-Goumard}, {Lenain},
  {Lohse}, {Marandon}, {Marcowith}, {Masbou}, {Maurin}, {McComb}, {Medina},
  {M{\'e}hault}, {Moulin}, {Naumann-Godo}, {de Naurois}, {Nedbal}, {Nekrassov},
  {Nicholas}, {Niemiec}, {Nolan}, {Ohm}, {Olive}, {Orford}, {Ostrowski},
  {Panter}, {Pedaletti}, {Pelletier}, {Petrucci}, {Pita}, {P{\"u}hlhofer},
  {Punch}, {Quirrenbach}, {Raubenheimer}, {Raue}, {Rayner}, {Reimer}, {Renaud},
  {de los Reyes}, {Rieger}, {Ripken}, {Rob}, {Rosier-Lees}, {Rowell}, {Rudak},
  {Rulten}, {Ruppel}, {Ryde}, {Sahakian}, {Santangelo}, {Schlickeiser},
  {Sch{\"o}ck}, {Sch{\"o}nwald}, {Schwanke}, {Schwarzburg}, {Schwemmer},
  {Shalchi}, {Sushch}, {Sikora}, {Skilton}, {Sol}, {Stawarz}, {Steenkamp},
  {Stegmann}, {Stinzing}, {Superina}, {Szostek}, {Tam}, {Tavernet}, {Terrier},
  {Tibolla}, {Tluczykont}, {van Eldik}, {Vasileiadis}, {Venter}, {Venter},
  {Vialle}, {Vincent}, {Vink}, {Vivier}, {V{\"o}lk}, {Volpe}, {Vorobiov},
  {Wagner}, {Ward}, {Zdziarski}, \& {Zech}}]{HESSsn1006}
{Aharonian}, F., {Akhperjanian}, A.~G., {Anton}, G., {et~al.} 2010, ArXiv
  e-prints

\bibitem[{{Blasi} {et~al.}(2005){Blasi}, {Gabici}, \& {Vannoni}}]{blasi2005}
{Blasi}, P., {Gabici}, S., \& {Vannoni}, G. 2005, \mnras, 361, 907

\bibitem[{{Cargill} \& {Papadopoulos}(1988)}]{Cargill1988}
{Cargill}, P.~J. \& {Papadopoulos}, K. 1988, \apjl, 329, L29

\bibitem[{{Chevalier}(1983)}]{Chevalier1983}
{Chevalier}, R.~A. 1983, \apj, 272, 765

\bibitem[{{Decourchelle} {et~al.}(2001){Decourchelle}, {Sauvageot}, {Audard},
  {Aschenbach}, {Sembay}, {Rothenflug}, {Ballet}, {Stadlbauer}, \&
  {West}}]{Decourchelle2001}
{Decourchelle}, A., {Sauvageot}, J.~L., {Audard}, M., {et~al.} 2001, \aap, 365,
  L218

\bibitem[{{Drury} {et~al.}(2009){Drury}, {Aharonian}, {Malyshev}, \&
  {Gabici}}]{Drury2009}
{Drury}, L.~O., {Aharonian}, F.~A., {Malyshev}, D., \& {Gabici}, S. 2009, \aap,
  496, 1

\bibitem[{{Dubner} {et~al.}(2002){Dubner}, {Giacani}, {Goss}, {Green}, \&
  {Nyman}}]{Dubner2002}
{Dubner}, G.~M., {Giacani}, E.~B., {Goss}, W.~M., {Green}, A.~J., \& {Nyman},
  L.-{\AA}. 2002, \aap, 387, 1047

\bibitem[{{Ellison} {et~al.}(2005){Ellison}, {Decourchelle}, \&
  {Ballet}}]{Ellison2005}
{Ellison}, D.~C., {Decourchelle}, A., \& {Ballet}, J. 2005, \aap, 429, 569

\bibitem[{{Ellison} {et~al.}(2007){Ellison}, {Patnaude}, {Slane}, {Blasi}, \&
  {Gabici}}]{Ellison2007}
{Ellison}, D.~C., {Patnaude}, D.~J., {Slane}, P., {Blasi}, P., \& {Gabici}, S.
  2007, \apj, 661, 879

\bibitem[{{Ghavamian} {et~al.}(2007){Ghavamian}, {Laming}, \&
  {Rakowski}}]{Ghavamian2007}
{Ghavamian}, P., {Laming}, J.~M., \& {Rakowski}, C.~E. 2007, \apjl, 654, L69

\bibitem[{{Ghavamian} {et~al.}(2001){Ghavamian}, {Raymond}, {Smith}, \&
  {Hartigan}}]{Ghavamian2001}
{Ghavamian}, P., {Raymond}, J., {Smith}, R.~C., \& {Hartigan}, P. 2001, \apj,
  547, 995

\bibitem[{{Ghavamian} {et~al.}(2002){Ghavamian}, {Winkler}, {Raymond}, \&
  {Long}}]{Ghavamian2002}
{Ghavamian}, P., {Winkler}, P.~F., {Raymond}, J.~C., \& {Long}, K.~S. 2002,
  \apj, 572, 888

\bibitem[{{Helder} {et~al.}(2009){Helder}, {Vink}, {Bassa}, {Bamba}, {Bleeker},
  {Funk}, {Ghavamian}, {van der Heyden}, {Verbunt}, \& {Yamazaki}}]{Helder2008}
{Helder}, E.~A., {Vink}, J., {Bassa}, C.~G., {et~al.} 2009, Science, 325, 719

\bibitem[{{Katsuda} {et~al.}(2009){Katsuda}, {Petre}, {Long}, {Reynolds},
  {Winkler}, {Mori}, \& {Tsunemi}}]{Katsuda2009}
{Katsuda}, S., {Petre}, R., {Long}, K.~S., {et~al.} 2009, \apjl, 692, L105

\bibitem[{{Korreck} {et~al.}(2004){Korreck}, {Raymond}, {Zurbuchen}, \&
  {Ghavamian}}]{Korreck2004}
{Korreck}, K.~E., {Raymond}, J.~C., {Zurbuchen}, T.~H., \& {Ghavamian}, P.
  2004, \apj, 615, 280

\bibitem[{{Laming}(2000)}]{Laming2000}
{Laming}, J.~M. 2000, \apjs, 127, 409

\bibitem[{{Laming}(2001)}]{Laming2001}
{Laming}, J.~M. 2001, \apj, 563, 828

\bibitem[{{Laming}(2004)}]{Laming2004}
{Laming}, J.~M. 2004, \apj, 604, 874

\bibitem[{{Laming} \& {Lepri}(2007)}]{Laming2007}
{Laming}, J.~M. \& {Lepri}, S.~T. 2007, \apj, 660, 1642

\bibitem[{{Laming} {et~al.}(1996){Laming}, {Raymond}, {McLaughlin}, \&
  {Blair}}]{Laming1996}
{Laming}, J.~M., {Raymond}, J.~C., {McLaughlin}, B.~M., \& {Blair}, W.~P. 1996,
  \apj, 472, 267

\bibitem[{{Long} {et~al.}(2003){Long}, {Reynolds}, {Raymond}, {Winkler},
  {Dyer}, \& {Petre}}]{Long2003}
{Long}, K.~S., {Reynolds}, S.~P., {Raymond}, J.~C., {et~al.} 2003, \apj, 586,
  1162

\bibitem[{{Patnaude} {et~al.}(2009){Patnaude}, {Ellison}, \&
  {Slane}}]{Patnaude2009}
{Patnaude}, D.~J., {Ellison}, D.~C., \& {Slane}, P. 2009, \apj, 696, 1956

\bibitem[{{Rakowski}(2005)}]{rakowski2005}
{Rakowski}, C.~E. 2005, Advances in Space Research, 35, 1017

\bibitem[{{Rakowski} {et~al.}(2008){Rakowski}, {Laming}, \&
  {Ghavamian}}]{Rakowski2008}
{Rakowski}, C.~E., {Laming}, J.~M., \& {Ghavamian}, P. 2008, \apj, 684, 348

\bibitem[{{Raymond} {et~al.}(2007){Raymond}, {Korreck}, {Sedlacek}, {Blair},
  {Ghavamian}, \& {Sankrit}}]{Raymond2007}
{Raymond}, J.~C., {Korreck}, K.~E., {Sedlacek}, Q.~C., {et~al.} 2007, \apj,
  659, 1257

\bibitem[{{Rothenflug} {et~al.}(2004){Rothenflug}, {Ballet}, {Dubner},
  {Giacani}, {Decourchelle}, \& {Ferrando}}]{Rothenflug2004}
{Rothenflug}, R., {Ballet}, J., {Dubner}, G., {et~al.} 2004, \aap, 425, 121

\bibitem[{{Rybicki} \& {Lightman}(1986)}]{Rybicki-Lightman}
{Rybicki}, G.~B. \& {Lightman}, A.~P. 1986, {Radiative Processes in
  Astrophysics} (Radiative Processes in Astrophysics, by George B.~Rybicki,
  Alan P.~Lightman, pp.~400.~ISBN 0-471-82759-2.~Wiley-VCH , June 1986.)

\bibitem[{{Sharma} {et~al.}(2007){Sharma}, {Quataert}, {Hammett}, \&
  {Stone}}]{Sharma2007}
{Sharma}, P., {Quataert}, E., {Hammett}, G.~W., \& {Stone}, J.~M. 2007, \apj,
  667, 714

\bibitem[{{Smith} {et~al.}(1991){Smith}, {Kirshner}, {Blair}, \&
  {Winkler}}]{Smith1991}
{Smith}, R.~C., {Kirshner}, R.~P., {Blair}, W.~P., \& {Winkler}, P.~F. 1991,
  \apj, 375, 652

\bibitem[{{Spitzer}(1998)}]{Spitzer}
{Spitzer}, L. 1998, {Physical Processes in the Interstellar Medium} (Physical
  Processes in the Interstellar Medium, by Lyman Spitzer, pp.~335.~ISBN
  0-471-29335-0.~Wiley-VCH , May 1998.)

\bibitem[{{Vink}(2008)}]{Vink2008}
{Vink}, J. 2008, in American Institute of Physics Conference Series, Vol. 1085,
  American Institute of Physics Conference Series, ed. {F.~A.~Aharonian,
  W.~Hofmann, \& F.~Rieger}, 169--180

\bibitem[{{Vink} {et~al.}(2003){Vink}, {Laming}, {Gu}, {Rasmussen}, \&
  {Kaastra}}]{Vink2003}
{Vink}, J., {Laming}, J.~M., {Gu}, M.~F., {Rasmussen}, A., \& {Kaastra}, J.~S.
  2003, \apjl, 587, L31

\bibitem[{{Winkler} {et~al.}(2003){Winkler}, {Gupta}, \& {Long}}]{Winkler2003}
{Winkler}, P.~F., {Gupta}, G., \& {Long}, K.~S. 2003, \apj, 585, 324

\end{thebibliography}
\end{document}